\documentclass[
superscriptaddress,
reprint,
amsmath,amssymb,
aps, 
pra,
floatfix,
]{revtex4-2}
\usepackage[english]{babel}
\usepackage[
    colorlinks=true,
    frenchlinks=false,
    linkcolor=blue,
    anchorcolor=blue,
    citecolor=blue,
    filecolor=blue,
    urlcolor=blue,
    bookmarks=true,
    bookmarksopen=true,
    bookmarksnumbered=true,
    bookmarksopenlevel=1,
    plainpages=false,
    pdfpagelabels=true,
    draft=false
]{hyperref} 

\usepackage{amsmath}
\usepackage{amssymb}
\usepackage{graphicx} 
\usepackage[separate-uncertainty=true, multi-part-units=single]{siunitx} 
\usepackage{xcolor} 
\DeclareSIUnit\dBm{dBm}
\DeclareSIUnit\bar{bar}
\usepackage[mathlines]{lineno} 
\usepackage{bbold}
\usepackage{braket}
\usepackage{comment}

\begin{document}


\title{Practical quantum teleportation with finite-energy codebooks}


\author{W.~K.~Yam}
\email{WunKwan.Yam@wmi.badw.de}
\affiliation{Walther-Mei{\ss}ner-Institut, Bayerische Akademie der Wissenschaften, 85748 Garching, Germany}
\affiliation{School of Natural Sciences, Technical University of Munich, 85748 Garching, Germany}

\author{M.~Renger}
\affiliation{Walther-Mei{\ss}ner-Institut, Bayerische Akademie der Wissenschaften, 85748 Garching, Germany}

\author{S.~Gandorfer}
\affiliation{Walther-Mei{\ss}ner-Institut, Bayerische Akademie der Wissenschaften, 85748 Garching, Germany}
\affiliation{School of Natural Sciences, Technical University of Munich, 85748 Garching, Germany}

\author{R.~Gross}
\affiliation{Walther-Mei{\ss}ner-Institut, Bayerische Akademie der Wissenschaften, 85748 Garching, Germany}
\affiliation{School of Natural Sciences, Technical University of Munich, 85748 Garching, Germany}
\affiliation{Munich Center for Quantum Science and Technology (MCQST), 80799 Munich, Germany}

\author{K.~G.~Fedorov}
\email{Kirill.Fedorov@wmi.badw.de}
\affiliation{Walther-Mei{\ss}ner-Institut, Bayerische Akademie der Wissenschaften, 85748 Garching, Germany}
\affiliation{School of Natural Sciences, Technical University of Munich, 85748 Garching, Germany}
\affiliation{Munich Center for Quantum Science and Technology (MCQST), 80799 Munich, Germany}


\begin{abstract}
Quantum communication exploits non-classical correlations to achieve efficient and unconditionally secure exchange of information. In particular, the quantum teleportation protocol allows for a deterministic and secure transfer of unknown quantum states by using pre-shared quantum entanglement and classical feedforward communication. Quantum teleportation in the microwave regime provides an important tool for high-fidelity remote quantum operations, enabling distributed quantum computing with superconducting circuits and potentially facilitating short-range, open-air microwave quantum communication. In this context, we consider practical application scenarios for the microwave analog quantum teleportation protocol based on continuous-variable states. We theoretically analyze the effect of feedforward losses and noise on teleportation fidelities of coherent states and show that these imperfections can be fully corrected by an appropriate feedforward gain. Furthermore, we consider quantum teleportation with finite-size codebooks and derive modified no-cloning thresholds as a function of the codebook configuration. Finally, we analyze the security of quantum teleportation under public channel attacks and demonstrate that the corresponding secure fidelity thresholds may drastically differ from the conventional no-cloning values. Our results contribute to the general development of quantum communication protocols and, in particular, illustrate the feasibility of using quantum teleportation in realistic microwave networks for robust and unconditionally secure communication.
\end{abstract}

\maketitle


\section{Introduction}
\label{Sec:Introduction}

Quantum communication utilizes the laws of quantum physics to outperform classical communication protocols in terms of efficiency and security. One of the fundamental quantum communication protocols is quantum teleportation, where an unknown quantum state is transferred between two distant parties using a shared entanglement resource and a classical feedforward~\cite{Bennett1993}. Originally, quantum teleportation was demonstrated at optical carrier frequencies in both discrete-variable~\cite{Bouwmeester1997} and continuous-variable (CV) regimes~\cite{Furusawa1998}. Later, CV quantum teleportation between distant superconducting nodes has also been implemented at microwave frequencies~\cite{Fedorov2021,Yam2025}. Here, CV quantum communication is especially attractive due to its high bit rates~\cite{Pirandola2015} and a comparatively robust experimental implementation. With the advancement of quantum information processors based on superconducting circuits~\cite{Google2025}, there is growing interest in microwave quantum communication, which can interconnect local superconducting quantum processors and solve the scalability challenge. In the context of such practical applications, we investigate the performance of microwave quantum teleportation under realistic conditions. In particular, we address the susceptibility to losses and noise, as well as the utilization of finite-energy, truncated codebooks, as opposed to typical Gaussian codebooks with infinite tails. While our research is motivated by microwave quantum communication, its results are applicable to quantum communication protocols in general, irrespective of the carrier frequencies and experimental platforms.
\begin{figure*}[t]
	\centering
	\includegraphics[width=\linewidth]{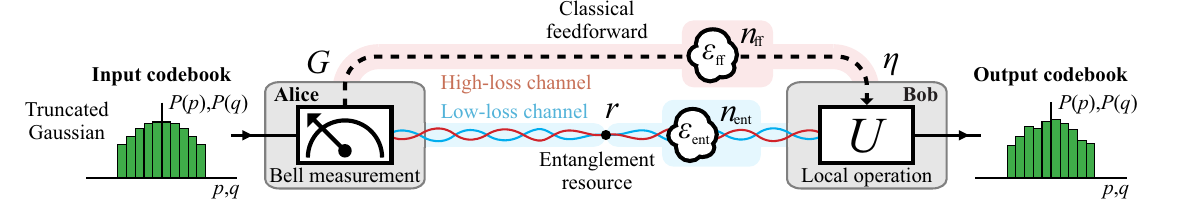}
	\caption{Schematic of a practical quantum teleportation protocol. A two-mode squeezed state is shared between Alice and Bob through a low-loss quantum channel. Alice performs a Bell-type measurement operation on an input code state superimposed with her part of the entangled state. Then, the measurement result is sent to Bob through a lossy and noisy feedforward channel. Finally, Bob performs a local unitary operation on his part of the entangled state according to the received feedforward signal in order to reconstruct the input state at his location. The input states are drawn from an ensemble of code states. The latter forms a certain codebook, which is represented by a truncated Gaussian distribution in this particular scheme.}
	\label{Fig1}
\end{figure*}

This paper is structured as follows: \hyperref[Section:analog_CVQT]{Section\,II} reviews the quantum teleportation protocol with coherent states and investigates it under realistic considerations. \hyperref[Section:TruncatedCodebooks]{Section\,III} introduces quantum communication using finite-energy codebooks and derives corresponding no-cloning fidelity thresholds, which provide the connection to unconditionally secure communication. \hyperref[Section:PublicEavesdropper]{Section\,IV} examines the security of quantum teleportation under a specific scenario of attack on the public feedforward channel. \hyperref[Section:Conclusion]{Section\,V} concludes this work and discusses the main implications of our results.

\section{Continuous-variable quantum teleportation of coherent states}
\label{Section:analog_CVQT}

Quantum teleportation enables the disembodied transfer of unknown quantum states between two spatially separated communication parties, conventionally named Alice and Bob~\cite{Bennett1993}. In a general quantum teleportation experiment, Alice utilizes a pre-shared entanglement resource and a classical feedforward channel to teleport an unknown input quantum state to Bob, as illustrated in \hyperref[Fig1]{Figure\,1}. The protocol is initiated by distributing quantum entanglement between Alice and Bob, followed by Alice entangling the input state with her part of the shared resource state. Then, she performs a Bell-type measurement operation on her resulting bipartite state, which contains correlations between both the input and resource states. Her measurement outcome, referred to as the feedforward, is transmitted to Bob through a corresponding lossy and noisy feedforward channel. Finally, Bob performs a local unitary operation on his part of the shared resource state, conditioned on the received feedforward signal, which allows him to obtain the input state at his location without actually propagating it through any communication channel.

Here, we focus on an analog CV quantum teleportation scheme with an input codebook formed by coherent states~\cite{DiCandia2015}. In this scenario, the entanglement resource is a two-mode squeezed (TMS) vacuum state. The Bell-type measurement operation by Alice consists of conjugate quadrature measurements. The unitary operation by Bob is a displacement operation in phase-space defined by the feedforward signal. Every input quantum state to be teleported is assumed to be a pure coherent state with a complex displacement amplitude $\alpha$. The TMS entanglement resource is produced by superimposing two orthogonally squeezed states with a squeezing level $S$ at a balanced beam splitter. For microwave applications, coherent signals are generated by room-temperature commercial vector microwave sources, while the individual squeezed states are produced by using superconducting Josephson parametric amplifiers (JPAs)~\cite{Fedorov2018}. The Bell-type measurement can be implemented by using a SU(1,1) nonlinear interferometer, which consists of two balanced beam splitters that allow two JPAs to amplify orthogonal field quadratures with a measurement gain $G$ along two parallel paths before combining them to produce the feedforward signal~\cite{Fedorov2021,Kronowetter2023}. One can show that such an operation coincides with a projective quadrature measurement in the limit of a high JPA degenerate gain $G \gg 1$~\cite{Fedorov2021}. Such an approach is also known as a measurement-device-free scheme~\cite{Yamashima2025}. The resulting analog feedforward signal is transmitted through an imperfect communication channel to Bob, who performs the displacement operation by employing a highly asymmetric beam splitter. The latter couples the feedforward signal to Bob's part of the entangled resource state with coupling strength $\eta \ll 1$ and implements the displacement operation~\cite{Fedorov2016}. A corresponding full model of the coherent state teleportation protocol is provided in \hyperref[Appendix:CoherentStateTeleportation]{Appendix\,A}.

For an arbitrary coherent state communication protocol, the information content is represented by the complex displacement amplitude of the input state, which is drawn from a specific codebook distribution. When utilizing the analog teleportation protocol, information about the input state displacement is carried by the feedforward signal, and the shared TMS entanglement serves as both an encoder and decoder for this information. Alice's part of the TMS state adds (or encodes) a ``quantum" noise variance to the input state displacement, which is sent over the feedforward channel and can later be removed (or decoded) by quantum-coherent interference with Bob's part of the TMS state. In the limit of infinite measurement gain, $G \to \infty$, we perform a perfect projection measurement to generate a completely classical feedforward signal which recovers the scenario of a digital feedforward used in other CV teleportation experiments~\cite{Furusawa1998}. In the limit of infinite resource squeezing, $S \to \infty$, we have perfect, EPR-type, entanglement and the TMS variance covers the entire signal space, which recovers the scenario of digital quantum teleportation protocols with maximally entangled Bell states. Due to this infinitely large TMS noise variance, the encoding step produces an infinitely noisy state, which is analogous to a one-time pad encryption~\cite{Schumacher2006,Brandao2012}. Thus, the security of the public feedforward signal in a general teleportation protocol is derived from the strength of its pre-shared entanglement resource. We show in \hyperref[Section:PublicEavesdropper]{Section\,IV} that a finite amount of information can be leaked to a potential eavesdropper via the public feedforward channel when the entanglement resource is not perfect.

Since the analog coherent state teleportation protocol involves only Gaussian operations, we can apply the covariance matrix formalism in the phase space spanned by electromagnetic field quadratures $p,q$~\cite{Weedbrook2012}. First- and second-order quadrature moments are sufficient to fully describe arbitrary Gaussian states. In the following, we use the vacuum variance definition of $1/4$ per field quadrature. The performance of a quantum teleportation procedure can be estimated by measuring a statistical overlap between the input and output states using the Uhlmann fidelity~\cite{Weedbrook2012,Scutaru1998}
\begin{equation}\label{Eq:fidelity}
	F(\alpha_1, V_1, \alpha_2, V_2) = \frac{1}{2} \frac{\exp\left( -\frac{1}{2} d^\intercal (V_1+V_2)^{-1} d \right)}{\sqrt{\Lambda + \Delta} - \sqrt{\Delta}},
\end{equation}
with
\begin{align}
	\Lambda &= \det(V_1+V_2), \\
	\Delta &= 16 (\det V_1 - 1/16) (\det V_2 - 1/16), \\
	d &= \alpha_1 - \alpha_2 \; .
\end{align}
Here, $\alpha_1$, $V_1$ and $\alpha_2$, $V_2$ are the complex displacement vectors and covariance matrices of states 1 and 2, respectively. To characterize communication using a specific codebook of coherent states, we average over the fidelities for the individual input states weighted by their corresponding probabilities in the codebook. In particular, an infinitely large codebook of coherent states gives the asymptotic classical threshold fidelity $F_\mathrm{cl} = 1/2$, which can only be exceeded with the use of quantum entanglement~\cite{Braunstein2001b}. Furthermore, for this choice of codebook, unconditionally secure communication is achieved if an average teleportation fidelity exceeds the asymptotic no-cloning threshold $F_\mathrm{nc} = 2/3$~\cite{Grosshans2001}. These threshold fidelities provide important and helpful benchmarks for evaluating various quantum communication schemes, including quantum teleportation. Nonetheless, in practical scenarios, using such infinite codebooks is impossible due to the infinite energy that is required to generate infinitely displaced input coherent states. This means that one needs to consider finite-energy codebooks with limited displacement amplitudes, which may lead to considerably higher no-cloning thresholds, $F > 2/3$. We investigate this important aspect of practical quantum teleportation in \hyperref[Section:TruncatedCodebooks]{Section\,III}.

\subsection{Ideal analog quantum teleportation}

First, we analyze an ideal analog CV quantum teleportation protocol, where all components are assumed to be lossless and noiseless. We further assume that the input states are perfectly coherent states, which correspond to the following displacement vector and covariance matrix at Alice
\begin{align}
	d_\mathrm{Alice} &= (\Re(\alpha), \Im(\alpha))^\intercal,
\\
	V_\mathrm{Alice} &= \frac{1}{4} \mathbb{1}_2,
\end{align}
while the output state displacement vector and covariance matrix at Bob are given by
\begin{align}
	d_\mathrm{Bob} &= \frac{\sqrt{G_+\eta}}{2} d_\mathrm{Alice}, \\
    \begin{split}
    V_\mathrm{Bob} &= \frac{1}{16} \biggl\{ G_+\eta + [G_-\eta + 4(1 - \eta)] \cosh(2r) \\
    &\qquad\qquad- 4\sqrt{G_-\eta (1-\eta)} \sinh(2r) \biggr\} \mathbb{1}_2,
    \end{split}
\end{align}
where $G_\pm = ( G^{1/2} \pm G^{-1/2} )^2$, $r = S/20\ln(10)$ is the squeezing factor, and $\mathbb{1}_2$ is the $2 \times 2$ identity matrix. For genuine quantum teleportation, a sufficiently large gain $G \gg 1$ is needed in order to have strongly projective measurements, whereby $G_\pm \to G$. Full formalism for the analog quantum teleportation protocol is provided in \hyperref[Appendix:CoherentStateTeleportation]{Appendix\,A}.

We see that an optimal output state displacement is achieved for the measurement gain $G = 4/\eta$, which corresponds to a full compensation of path losses due to the two beam splitters in the SU(1,1) measurement interferometer and the coupling strength $\eta$~\cite{DiCandia2015}. Satisfying this displacement-matching condition $G\eta = 4$ allows the output state displacement to match the input state displacement. It also balances the quantum noise variance in the feedforward signal with that in Bob's entangled state, thereby optimizing the interference between the TMS correlations of the two paths. In the following analysis, we will always assume that the displacement-matching condition is fulfilled. While for small input state displacements it is advantageous to use lower $G$ values to reduce the output state variance, this advantage of gain-tuning disappears for higher displacement photon numbers $|\alpha|^2 \gtrsim 10$. Nevertheless, we consider this effect for teleportation with energetic constraints in \hyperref[Section:TruncatedCodebooks]{Section\,III}.

\subsection{Quantum teleportation over thermal communication channels}

\begin{figure*}[t]
	\centering
	\includegraphics[width=\linewidth]{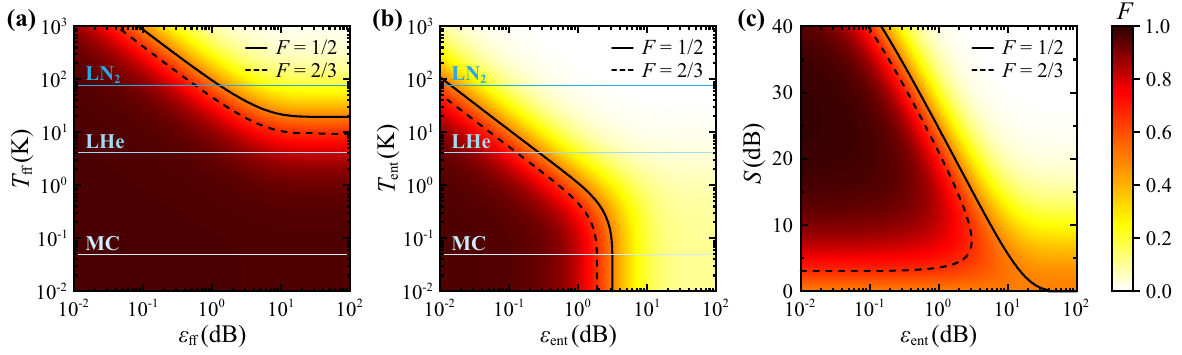}
	\caption{Microwave teleportation fidelities through thermal communication channels at the carrier frequency of $\omega/2\pi = \SI{5}{\giga \hertz}$ and for a displacement photon number of $|\alpha|^2 = \SI{10}{}$, squeezing level of $S = \SI{10}{\decibel}$, and feedforward coupling of $\eta = \SI{-20}{\decibel}$. Horizontal lines indicate characteristic mixing chamber, liquid helium, and liquid nitrogen temperatures of $\SI{50}{\milli\kelvin}$, $\SI{4.2}{\kelvin}$, and $\SI{77}{\kelvin}$, respectively. Black solid and dashed lines indicate the asymptotic classical limit $F_\mathrm{cl} = 1/2$ and asymptotic no-cloning limit $F_\mathrm{nc} = 2/3$, respectively. (a) Teleportation fidelity $F$ as a function of the feedforward channel losses $\varepsilon_\mathrm{ff}$ and noise temperature $T_\mathrm{ff}$, where $\varepsilon_\mathrm{ent} = 0$. (b) Teleportation fidelity $F$ as a function of the entanglement distribution channel losses $\varepsilon_\mathrm{ent}$ and noise temperature $T_\mathrm{ent}$, where $\varepsilon_\mathrm{ff} = 0$. (c) Teleportation fidelity $F$ as a function of the entanglement distribution channel losses $\varepsilon_\mathrm{ent}$ and resource squeezing level $S$, where $\varepsilon_\mathrm{ff} = 0$.}
	\label{Fig2}
\end{figure*}
Microwave quantum communication is especially susceptible to ambient thermal noise due to the relatively low energies of microwave photons compared to the thermal noise environment typically found in experiments. Therefore, it is important to consider the impact of finite channel temperatures and explore possible mitigation strategies. In this context, we consider the analog CV quantum teleportation scheme in the presence of realistic imperfections in communication channels, as illustrated in \hyperref[Fig1]{Figure\,1}. We analyze the impact of power losses in the feedforward channel $\varepsilon_\mathrm{ff}$ and in the entanglement distribution channel $\varepsilon_\mathrm{ent}$, which are coupled to local bosonic baths with temperatures $T_\mathrm{ff}$ and $T_\mathrm{ent}$, respectively. In particular, we are only considering path losses for the signals propagating towards Bob. Apart from these imperfections, we continue to treat the rest of the quantum teleportation protocol as lossless and noiseless.

We find that the feedforward losses $\varepsilon_\mathrm{ff}$ act as a simple renormalization of the displacement-matching condition, which is then given by $G\eta (1-\varepsilon_{\mathrm{ff}}) = 4$. The non-zero entanglement distribution losses $\varepsilon_\mathrm{ent}$ may also add an additional term to this condition, leading to $G\eta(1-\varepsilon_\mathrm{ff})/(1-\varepsilon_\mathrm{ent}) = 4$, which compensates for the imbalanced TMS correlations. However, at experimentally relevant TMS squeezing levels $S \lesssim \SI{20}{\decibel}$, the teleportation fidelity degrades more quickly with the displacement mismatch than with the TMS imbalance. Hence, we only use the condition $G\eta(1-\varepsilon_\mathrm{ff}) = 4$. This behavior can be understood from equation\,(\ref{Eq:fidelity}), where for a fixed squeezing level $S$, the fidelity decays exponentially with displacement mismatch and polynomially with TMS imbalance.

Now, using the assumptions $G \gg 1$ and $G\eta(1-\varepsilon_\mathrm{ff}) = 4$, we can derive the output state displacement vector and covariance matrix for the lossy transmission channels as
\begin{align}
	d_\mathrm{Bob} &= d_\mathrm{Alice}, \\
    \begin{split}
	V_\mathrm{Bob} &= v_\mathrm{out} \mathbb{1}_2 = \frac{1}{4} \biggl\{ \left[ 1 + \varepsilon_\mathrm{ent} W_\mathrm{ent} + \eta \varepsilon_\mathrm{ff} W_\mathrm{ff} \right] \\
    &\quad\qquad\qquad\qquad+ \left[ 2-\varepsilon_\mathrm{ent} \right] \cosh(2r) \\
    &\quad\qquad\qquad\qquad- \left[ 2\sqrt{ 1-\varepsilon_\mathrm{ent} } \right] \sinh(2r) \biggr\} \mathbb{1}_2.
    \end{split}\label{Eq:outputCovariance}
\end{align}
To simplify notation, we define $W_\mathrm{ff} = 2n_\mathrm{ff}+1$ and $W_\mathrm{ent} = 2n_\mathrm{ent}+1$ as noise terms from the ambient thermal photon populations, $n_\mathrm{ff}$ and $n_\mathrm{ent}$, in the feedforward and entanglement distribution channels, respectively. Note that for $G \gg 1$ and fixed $\varepsilon_\mathrm{ff}$, we also have $\eta \ll 1$, but still keep the $\eta\varepsilon_\mathrm{ff}W_\mathrm{ff}$ term because $W_\mathrm{ff}$ can be large.

Assuming a pure input coherent state and $\varepsilon_\mathrm{ent} = 0$, the teleported state fidelity can be expressed as
\begin{equation}\label{Eq:FidelityNoisyFF}
	F = \frac{2}{1 + 4 v_\mathrm{out}} = \frac{1}{1 + e^{-2r} + \eta\varepsilon_\mathrm{ff} W_\mathrm{ff}/2}.
\end{equation}
We observe that analog quantum teleportation implements an error correction on the feedforward channel, as the thermal noise $W_\mathrm{ff}$ can be completely suppressed by a sufficiently small coupling strength $\eta \ll 1$. Furthermore, the feedforward losses $\varepsilon_\mathrm{ff}$ can be compensated by a corresponding gain $G = 4/[\eta(1-\varepsilon_{\mathrm{ff}})]$. This behavior is also translated to the scenario of a digital feedforward, corresponding to the limit $\eta \to 0$ and $G \to \infty$.

Next, we numerically simulate the analog teleportation protocol in a wide range of transmission channel losses and noise by choosing a characteristic carrier frequency of $\omega/2\pi = \SI{5}{\giga\hertz}$, and setting the experimentally feasible values of $|\alpha|^2 = \SI{10}{}$, $S = \SI{10}{\decibel}$, and $\eta = \SI{-20}{\decibel}$. The measurement gain $G$ is adjusted to fulfill the displacement-matching condition $G = 4/[\eta(1-\varepsilon_{\mathrm{ff}})]$ at all times. \hyperref[Fig2]{Figure\,2}(a) shows the teleportation fidelities for the thermal feedforward channel. For low $\varepsilon_\mathrm{ff}$, the fidelity $F$ degrades approximately proportionally to the product $\varepsilon_\mathrm{ff} T$, which scales approximately linearly with the number of added noise photons $n_\mathrm{ff}$. For high $\varepsilon_\mathrm{ff}$, there is a plateau in $F$ because of the correspondingly large $G$. Even in the high-loss limit, it is always possible to obtain teleportation fidelities beyond the no-cloning threshold $F_\mathrm{nc} = 2/3$ at ambient temperatures of $T_\mathrm{ff} \simeq \SI{10}{\kelvin}$. In microwave experiments, this means that we are practically limited by the critical temperature of superconducting transmission lines used for communication, which is $T_\mathrm{c} = \SI{9.8}{\kelvin}$ for conventional NbTi coaxial cables.

\begin{figure*}[t]
	\centering
	\includegraphics[width=\linewidth]{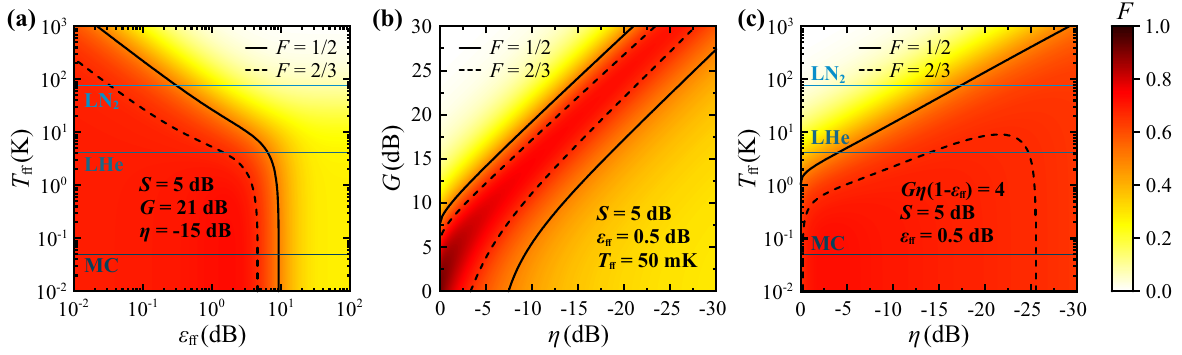}
	\caption{Teleportation fidelity as a function of various realistic experimental parameters. All experimental losses and noise sources are considered, based on the parameters in Ref.\,\cite{Yam2025}. Horizontal lines indicate characteristic mixing chamber, liquid helium, and liquid nitrogen temperatures of $\SI{50}{\milli\kelvin}$, $\SI{4.2}{\kelvin}$, and $\SI{77}{\kelvin}$, respectively. Black solid and dashed lines indicate the asymptotic classical limit $F_\mathrm{cl}=1/2$ and asymptotic no-cloning limit $F_\mathrm{nc}=2/3$, respectively. (a) Teleportation fidelity $F$ as a function of feedforward channel loss $\varepsilon_\mathrm{ff}$ and noise temperature $T_\mathrm{ff}$. The fidelity maintains a quantum advantage up to $\varepsilon_\mathrm{ff} \simeq \SI{7}{\decibel}$ and liquid helium temperatures. (b) Teleportation fidelity $F$ as a function of feedforward coupling strength $\eta$ and measurement gain $G$. Fidelity maxima occur near the displacement-matching condition $G\eta(1-\varepsilon_\mathrm{ff}) = 4$. (c) Teleportation fidelity $F$ as a function of feedforward coupling strength $\eta$ and noise temperature $T_\mathrm{ff}$, where the displacement-matching condition is fulfilled. The fidelity maintains quantum advantage up to room temperature at $\eta \simeq \SI{-24}{\decibel}$.}
	\label{Fig3}
\end{figure*}

We note, however, that the teleportation fidelity is inevitably reduced for $\varepsilon_\mathrm{ent} > 0$. Due to the no-cloning theorem, neither lost signal nor added noise in the entanglement distribution channel can be compensated for by a simple amplification or attenuation. This effect is shown in \hyperref[Fig2]{Figure\,2}(b), where we see that quantum teleportation is always susceptible to imperfections in the entanglement distribution channel. Comparing \hyperref[Fig2]{Figures\,2}(a) and \hyperref[Fig2]{2}(b) reveals that teleportation over thermal communication channels is limited by the entanglement distribution losses. Furthermore, a finite value of $\varepsilon_\mathrm{ent}$ induces an imbalance between the two paths of the TMS resource state, which constrains the optimal squeezing level $S$ (or squeezing factor $r$) by
\begin{equation}\label{Eq:optimal_squeezing}
	\varepsilon_\mathrm{ent} \cosh^2(2r) = 1.
\end{equation}
This is seen in \hyperref[Fig2]{Figure\,2}(c), where $\varepsilon_\mathrm{ent}$ limits the benefit of higher squeezing levels $S$. For a realistic experimental setup with $\varepsilon_\mathrm{ent} \simeq \SI{1}{\decibel}$, the optimal squeezing level is around $S = \SI{8}{\decibel}$. Generally, for any protocol that utilizes the TMS entanglement resources, asymmetric path losses always constrain the benefit of high squeezing.

\subsection{Practical quantum teleportation of microwave coherent states}

Passive components and amplifiers also contribute to losses and noise in experimental implementations. Most of these imperfections cannot be compensated for by the measurement gain $G$ because they mainly occur before the Bell-type measurement operation, during the amplification process, or in the entanglement distribution channel. Here, we include all experimental imperfections based on parameters from Ref.\,\cite{Yam2025} and obtain fidelity values in good agreement with the experimental data. In particular, we have a carrier frequency of $\SI{5.35}{\giga\hertz}$, coherent states with average displacement of $\SI{1.3}{}$ photons, a TMS squeezing level of $\SI{5}{\decibel}$, total losses of $\SI{1.6}{\decibel}$, and a noise bath temperature of $\SI{50}{\milli\kelvin}$. Here, the optimal squeezing level $S$ is determined by the quantum efficiencies of the JPAs used to generate the TMS resource state.

\hyperref[Fig3]{Figure\,3}(a) shows the teleportation fidelity as a function of the feedforward channel losses $\varepsilon_{\mathrm{ff}}$ and the feedforward noise bath temperature $T_{\mathrm{ff}}$. In contrast to an idealized scenario in \hyperref[Fig2]{Figure\,2}(a), experimental parameters $\eta = \SI{-15}{\decibel}$ and $G = \SI{21}{\decibel}$ are fixed, leading to a non-ideal displacement matching $G\eta(1-\varepsilon_\mathrm{ff}) \ne 4$. In practice, the feedforward losses may not be precisely known, and it can be difficult to exactly meet the displacement-matching condition. We observe that the realistic teleportation fidelities depend on feedforward imperfections in a qualitatively similar way as in the idealized case, but the fidelity values are lower, mainly due to gain-dependent JPA noise in the entanglement generation and Bell-type measurement procedures. Nevertheless, fidelities above $F_\mathrm{cl} = 1/2$ can still be attained at large losses $\varepsilon_\mathrm{ff} \simeq \SI{6}{\decibel}$ and up to liquid helium temperatures, i.e., up to temperatures well above the $\SI{10}{\milli\kelvin}$ base temperature of modern dilution refrigerators.

\begin{figure*}[t]
	\centering
	\includegraphics[width=\linewidth]{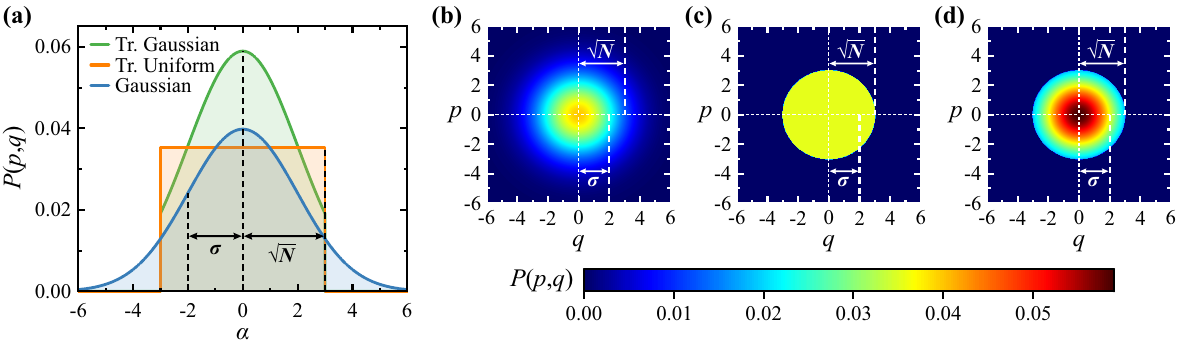}
	\caption{Probability distribution cross-sections of Gaussian, truncated uniform, and truncated Gaussian codebooks as a function of the complex displacement amplitude $\alpha$. The truncated uniform codebook is described by a cutoff photon number $N$ and is normalized to unit probability. The truncated Gaussian codebook is described by a Gaussian codebook with variance $\sigma^2$ and a cutoff photon number $N$, rescaled for normalization. Probability distribution functions in the phase space spanned by field quadratures ${p,q}$ for (b) Gaussian, (c) truncated uniform, and (d) truncated Gaussian codebooks.}
	\label{Fig4:codebooks}
\end{figure*}

In order to suppress the feedforward noise with a sufficiently weak coupling strength $\eta$, the measurement gain $G$ must be increased to compensate for the total attenuation. A realistic dependence of the teleportation fidelity on $\eta$ and $G$ is shown in \hyperref[Fig3]{Figure\,3}(b), where we see that fidelity maxima occur near the displacement-matching condition $G\eta(1-\varepsilon_{\mathrm{ff}}) = 4$. Moreover, when the displacement-matching condition is fulfilled, teleportation fidelities can exceed $F_\mathrm{cl} = 1/2$ even up to room temperatures at a coupling strength of $\eta \simeq \SI{-24}{\decibel}$, as shown in \hyperref[Fig3]{Figure\,3}(c). Although the value of $\eta$ can, in principle, be arbitrarily reduced by adding sufficiently large cold attenuation, a crucial constraint arises from the maximum JPA gain. As an example, narrow-band quantum-limited JPAs can reach degenerate gain values of up to $\SI{40}{\decibel}$, but are eventually limited by compression effects and suffer from the degradation of output state purity due to gain-dependent noise~\cite{Renger2021}. A practical challenge is finding an optimal balance between the JPA degenerate gain and noise to obtain the highest quantum teleportation fidelities.

\section{No-cloning threshold and truncated codebook}
\label{Section:TruncatedCodebooks}

In order for Alice to send information to Bob via a quantum communication protocol, the information-carrying quantum states must be drawn from a certain codebook distribution. During communication, a potential eavesdropper, Eve, can try to siphon this information. For quantum communication with coherent states, the optimal attack by Eve is to intercept and clone the transmitted code state, such that one copy is kept by Eve and the other copy is sent to Bob~\cite{Scarani2005,Cerf2000,Braunstein2001a}. Quantum communication protocols promise secure exchange of information due to the no-cloning theorem, which states that quantum states cannot be perfectly cloned. The unconditional, or information-theoretic, communication security is guaranteed if transmitted state fidelities, averaged over the entire codebook distribution, surpass the no-cloning threshold, since it will exceed the maximum fidelity physically available to any eavesdropper~\cite{Grosshans2002}.

In coherent state protocols, individual code states with complex displacement amplitudes $\alpha$ are drawn from a codebook with probability distribution $P(\alpha)$. In the case when $P(\alpha)$ uniformly covers the entire quadrature phase space, it can be shown that Eve's optimal (Gaussian) attack results in the no-cloning fidelity of $F_\mathrm{nc} = 2/3$~\cite{Grosshans2001}, which can be achieved by an entangling cloner attack~\cite{Braunstein2001a}. While non-Gaussian cloning can lead to a slightly higher cloning fidelity of $0.68$~\cite{Cerf2005}, such non-Gaussian cloners can be ruled out by checking the statistics of the feedforward signal. Hence, if Bob measures an average teleportation fidelity above $2/3$, he will know with certainty that he obtains more information than Eve, and consequently, his communication is unconditionally secure (assuming appropriate sifting and privacy amplification steps). However, any infinitely large codebook is experimentally impossible due to finite energy constraints. Practical codebook distributions must be limited in size. This leads to no-cloning thresholds larger than $F_\mathrm{nc} = 2/3$, because Eve can adjust her cloning process based on \emph{a priori} knowledge about the constrained codebook distribution. Therefore, it is important to analyze finite-energy codebooks (depicted in \hyperref[Fig4:codebooks]{Figure\,4}) and estimate the corresponding no-cloning fidelity thresholds.

\subsection{No-cloning threshold for a Gaussian codebook}

First, we review well-known results for Gaussian codebooks, for which an optimal cloning scheme is known~\cite{Cochrane2004}. Any Gaussian codebook (see \hyperref[Fig4:codebooks]{Figure\,4}(b)) can be described by the following probability distribution
\begin{equation}\label{Eq:gaussian}
    P(\alpha,\sigma) = \frac{1}{2\pi\sigma^2} e^{-|\alpha|^2 / 2\sigma^2},
\end{equation}
where $\alpha$ is the complex displacement amplitude of a code state and $\sigma^2$ is the codebook variance. The optimal average cloning fidelity for the Gaussian codebook is given by \cite{Cochrane2004}
\begin{align}
    F_\mathrm{nc}(\sigma) \le
    \begin{cases}
        \frac{4\sigma^2 + 2}{6\sigma^2 + 1} \qquad & \sigma^2 \ge \frac{1}{2} + \frac{1}{\sqrt{2}}, \\
        \frac{1}{(3 - 2\sqrt{2})\sigma^2 + 1} \qquad & \sigma^2 \leq \frac{1}{2} + \frac{1}{\sqrt{2}},
    \end{cases}
\end{align}
which defines the no-cloning threshold as a function of $\sigma^2$. In the limit of $\sigma \to \infty$, the Gaussian codebook becomes uniformly distributed and $F_\mathrm{nc}(\sigma) \to 2/3$.

\subsection{No-cloning threshold for truncated codebooks}

\begin{figure*}[t]
	\centering
	\includegraphics[width=\linewidth]{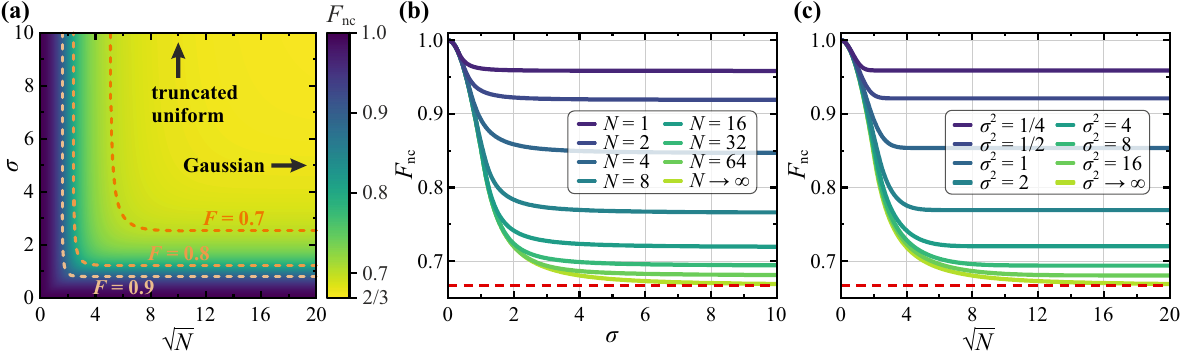}
	\caption{(a) No-cloning threshold for the truncated Gaussian codebook as a function of the cutoff photon number $N$ and codebook variance $\sigma^2$. Cross-sections of the no-cloning fidelity threshold for the truncated Gaussian codebook at various (b) $N$ and (c) $\sigma^2$. No-cloning thresholds for the Gaussian and truncated uniform codebooks are given by the plots where $N \to \infty$ and $\sigma^2 \to \infty$, respectively. Red dashed lines correspond to the asymptotic no-cloning threshold $F_\mathrm{nc} = 2/3$.}
	\label{Fig5:noCloning}
\end{figure*}
The Gaussian codebook has infinitely long tails that consist of individual coherent states with infinite energies, which are experimentally impossible to prepare. To solve this issue, we consider two finite-energy codebooks: truncated uniform and truncated Gaussian. The truncated uniform codebook (see \hyperref[Fig4:codebooks]{Figure\,4}(c)) is simply a uniform distribution limited to the region $|\alpha|^2 \le N$, and has a simple probability distribution
\begin{equation}\label{Eq:truncatedUniform}
	P(\alpha,N) = \frac{1}{\pi N} \Theta(N-|\alpha|^2),
\end{equation}
where $\Theta$ is the Heaviside step function. The truncated Gaussian codebook (see \hyperref[Fig4:codebooks]{Figure\,4}(d)) is constructed by truncating the Gaussian codebook with variance $\sigma^2$ at the cutoff photon number $N$, and then rescaling it to fulfill the normalization condition, resulting in
\begin{equation}\label{Eq:truncatedGaussian}
    P(\alpha,\sigma,N) = \frac{e^{-\frac{|\alpha|^2}{2\sigma^2}}}{2\pi\sigma^2 (1-e^{N/2\sigma^2})} \Theta(N - |\alpha|^2).
\end{equation}
From the truncated Gaussian distribution, we can recover the original Gaussian distribution by taking $N \to \infty$ or the truncated uniform distribution by taking $\sigma \to \infty$ and $\lim_{\sigma \to \infty} 2\sigma^2 (1-e^{-N/2\sigma^2}) = N$.

Now, we can calculate the no-cloning thresholds for these truncated codebooks. The fidelity obtainable with the entangling cloner attack on an arbitrary code state $\alpha$ is given by
\begin{equation}
	F(\alpha,A) = \frac{2}{1+A} \exp\bigg[ -\frac{2(1-\sqrt{A/2})^2 |\alpha|^2}{1+A} \bigg],
\end{equation}
where $A \ge 1$ is the amplification of the cloner. This expression can be derived by applying equation\,(\ref{Eq:fidelity}) to an input code state and its corresponding clone resulting from the entangling cloner process~\cite{Braunstein2001a}. We want to maximize the average fidelity
\begin{equation}
	\max_{A} \bar{F} = \max_{A} \left[ \int d\alpha~ F(\alpha,A) P(\alpha) \right],
\end{equation}
over the amplification $A$ and a certain codebook distribution $P(\alpha)$. This maximum cloning fidelity defines the no-cloning threshold. For the distributions in equations\,(\ref{Eq:truncatedUniform}) and (\ref{Eq:truncatedGaussian}), the calculation leads to transcendental functions, so there are no closed-form expressions. Hence, we numerically evaluate the maximum $\bar{F}$, which we denote as $F_\mathrm{nc}(\sigma,N)$ for the truncated Gaussian codebook distribution $P(\alpha,\sigma,N)$.

\hyperref[Fig5:noCloning]{Figure\,5} shows the no-cloning threshold $F_\mathrm{nc}(\sigma,N)$ for the truncated Gaussian codebook as a function of the codebook variance $\sigma^2$ and cutoff photon number $N$. We observe that $F_\mathrm{nc}(\sigma,N) \to 1$, when $N \to 0$ or $\sigma^2 \to 0$, which corresponds to an eavesdropper getting more information about a transmitted code state when the codebook is smaller. As can be seen in \hyperref[Fig5:noCloning]{Figure\,5}(b) and \hyperref[Fig5:noCloning]{5}(c), the truncated Gaussian no-cloning threshold approaches the Gaussian case for $N \to \infty$, and also the truncated uniform case for $\sigma^2 \to \infty$.

\section{Security under a public channel eavesdropper}
\label{Section:PublicEavesdropper}

The no-cloning thresholds discussed in \hyperref[Section:TruncatedCodebooks]{Section\,III} apply to any coherent state communication protocol, even when an eavesdropper Eve attacks all communication channels, including the entanglement distribution. However, in more realistic scenarios of quantum teleportation, it is stipulated that Eve may only have access to the public feedforward channel. The entanglement distribution channel is considered private, hence physically inaccessible to Eve. In this case, Eve can access less information about the teleported input states, resulting in lower threshold fidelity values that still guarantee security.

In the following, we define a particular secure fidelity $F_\mathrm{s}$, which corresponds to secure communication between Alice and Bob against an eavesdropper Eve on a public channel. The information shared between Alice and Bob is characterized by their mutual information $I(\mathrm{A}:\mathrm{B})$ and the information accessible to Eve is bounded from above by the Holevo quantity $\chi_\mathrm{E}$~\cite{Braunstein2005,Holevo1973}. Quantum teleportation is secure against this public channel eavesdropper as long as $I(\mathrm{A}:\mathrm{B}) > \chi_\mathrm{E}$. Since teleportation fidelity strictly increases with $I(\mathrm{A}:\mathrm{B})$ and strictly decreases with $\chi_\mathrm{E}$, we can find a minimum value of $F_\mathrm{s}$ that fulfills the condition $I(\mathrm{A}:\mathrm{B}) > \chi_\mathrm{E}$. We must stress here that if Eve can also attack the private entanglement distribution channel, then the secure fidelity $F_\mathrm{s}$ reduces to the no-cloning fidelity.

\subsection{Analog quantum teleportation with public channel eavesdropper}

To model a public channel eavesdropper, we consider the coupling of an environmental bath to the feedforward communication channel. This follows our analysis from \hyperref[Section:analog_CVQT]{Section\,II}, where we have considered finite $\varepsilon_\mathrm{ff}$ and zero losses elsewhere. In this way, Eve can perform an entangling cloner attack on the feedforward channel via losses $\varepsilon_\mathrm{ff}$~\cite{Braunstein2001a}. The bath noise photon number $n_\mathrm{ff}$ characterizes the power of Eve's entanglement TMS resource state. The full theory model and formalism are detailed in \hyperref[Appendix:CoherentStateTeleportation]{Appendix\,A}.

First, we analyze the case of the Gaussian codebook with variance $\sigma^2$. Since we are working with zero-mean Gaussian state ensembles, covariance matrices are enough to describe the information content. Covariance matrices of the quantum state ensembles held by Alice before the teleportation procedure, Eve after the entangling cloner attack, and Bob after the teleportation procedure are given by
\begin{align}
	V_\mathrm{Alice} &= \frac{1}{4} (4\sigma^2 + 1) \mathbb{1}_2, \\
	V_\mathrm{Eve} &= \frac{1}{4} \left[ \frac{ G (4\sigma^2+1) + G \cosh(2r)}{4} \right] \mathbb{1}_2. \\
	V_\mathrm{Bob} &= \frac{1}{4} \Bigl\{ (4 \sigma^2+1) + 2 e^{-2r} + \eta\varepsilon_\mathrm{ff} W_\mathrm{ff} \Bigr\} \mathbb{1}_2,
\end{align}
where the covariance matrices of individual coherent states are obtained by setting $\sigma = 0$. To simplify this analysis, we have again assumed an idealized teleportation protocol where Alice's measurement gain is sufficiently strong, $G \gg 1$, and that Bob's reconstruction fulfills the displacement-matching condition $G \eta (1-\varepsilon_\mathrm{ff}) = 4$. For fixed losses, $\varepsilon_\mathrm{ff} \in [0,1)$, these assumptions imply that Bob does not see any disturbance from the feedforward noise due to $\eta \ll 1$. Furthermore, since $G \gg 1$, the feedforward signal effectively becomes classical and Eve can obtain almost complete information about the feedforward measurement outcome.

\subsection{Mutual information and Holevo quantity}

Information shared between Alice and Bob is quantified by the mutual information, which can be calculated using~\cite{Braunstein2005}
\begin{equation}
	I(\mathrm{A}:\mathrm{B}) = \iint d\alpha d\beta~ P(\beta|\alpha) P(\alpha) \ln\left( \frac{ P(\beta|\alpha) }{ P(\beta) } \right),
\end{equation}
where $\alpha$ and $\beta$ are the complex displacement amplitudes of the input and teleported output states, respectively. The integration is done over the two-dimensional complex planes of $\alpha$ and $\beta$. Here, we utilize the input ensemble distribution $P(\alpha)$, the teleported ensemble distribution $P(\beta)$, and the conditional distribution $P(\beta|\alpha)$. The distribution $P(\alpha)$ is given by equations\,(\ref{Eq:gaussian}), (\ref{Eq:truncatedUniform}), and (\ref{Eq:truncatedGaussian}). The distributions $P(\beta)$ and $P(\beta|\alpha)$ are found using Bob's covariance matrix after the teleportation procedure (see \hyperref[Appendix:MutualInformationHolevo]{Appendix\,B}). The mutual information can also be expressed as~\cite{Shannon1948,Gyongyosi2018}
\begin{equation}
	I(\mathrm{A}:\mathrm{B}) \sim \ln(1+\mathrm{SNR}),
\end{equation}
where $\mathrm{SNR}$ is the signal-to-noise ratio at Bob's location. An upper bound for Eve's accessible information is given by the Holevo quantity~\cite{Holevo1973}
\begin{equation}\label{eq:holevo}
	\chi_\mathrm{E} = H(\hat{\rho}_\mathrm{E}) - \int d\alpha~ P(\alpha) H(\hat{\rho}_\alpha),
\end{equation}
where $H(\cdot)$ is the von Neumann entropy, $\hat{\rho}_\mathrm{E} = \int d\alpha~ P(\alpha) \hat{\rho}_\alpha$ is the state ensemble at Eve, and $\hat{\rho}_\alpha$ are the individual states at Eve corresponding to \emph{a priori} probability distribution $P(\alpha)$. These state entropies are found using Eve's covariance matrix after the entangling cloner attack. Methods to calculate the mutual information and Holevo quantity are provided in \hyperref[Appendix:MutualInformationHolevo]{Appendix\,B}.

\subsection{Secure fidelity threshold}

\begin{figure*}[t]
	\centering
	\includegraphics[width=\linewidth]{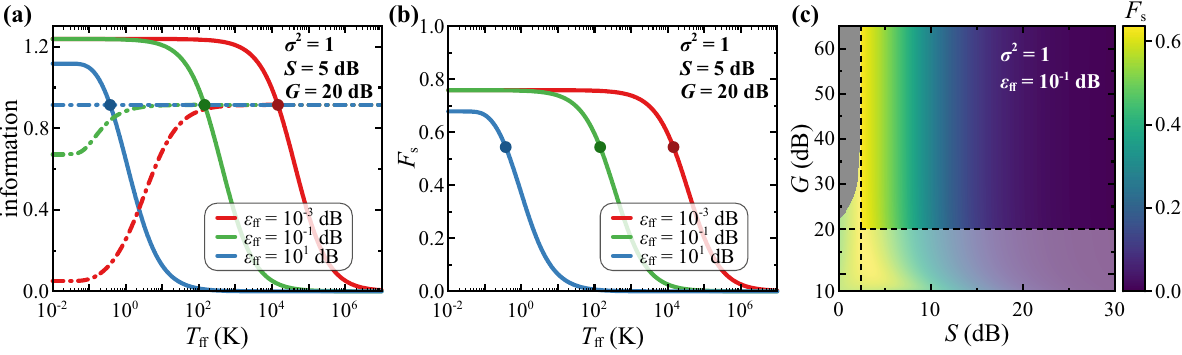}
	\caption{(a) Information content quantified by Alice's and Bob's mutual information $I(\mathrm{A}:\mathrm{B})$ (solid lines) and Eve's Holevo quantity $\chi_\mathrm{E}$ (dotted-dashed lines) for various feedforward losses $\varepsilon_\mathrm{ff}$ as a function of the feedforward channel temperature $T_\mathrm{ff}$. Dots indicate crossing points at $I(\mathrm{A}:\mathrm{B}) = \chi_\mathrm{E}$. (b) Teleportation fidelity for the corresponding parameters in panel (a). Dots correspond to the secure fidelity thresholds $F_\mathrm{s}$. (c) Secure fidelity $F_\mathrm{s}$ as a function of the resource squeezing level $S$ and measurement gain $G$ for a particular Gaussian codebook. Gray region denotes a parameter regime where security against a general feedforward attack cannot be achieved. Dashed lines denote the region where $S \ge \SI{2.39}{\decibel}$ and $G \ge \SI{20}{\decibel}$.}
	\label{Fig6:secureFidelity}
\end{figure*}

Now, we have to find parameter regimes where $I(\mathrm{A}:\mathrm{B}) > \chi_E$ and the corresponding $F_\mathrm{s}$ thresholds. \hyperref[Fig6:secureFidelity]{Figure\,6}(a) shows $I(\mathrm{A}:\mathrm{B})$ and $\chi_\mathrm{E}$ values for various operating parameters. We choose a particular codebook variance $\sigma^2=1$ without loss of generality, because the relative information content does not depend on codebook size. For low $\varepsilon_\mathrm{ff}$ and $T_\mathrm{ff}$, Eve has limited access to the feedforward signal, so $\chi_\mathrm{E}$ is smaller than $I(\mathrm{A}:\mathrm{B})$ and the communication between Alice and Bob is secure. As $\varepsilon_\mathrm{ff}$ and $T_\mathrm{ff}$ increase, $\chi_\mathrm{E}$ can eventually overtake $I(\mathrm{A}:\mathrm{B})$ and security is compromised. The teleportation fidelity at this crossing point (denoted by colored dots in \hyperref[Fig6:secureFidelity]{Figure\,6}(a) and \hyperref[Fig6:secureFidelity]{6}(b)) is the $F_\mathrm{s}$ fidelity threshold for which quantum teleportation is secure against any public channel eavesdropper. We observe that for any nonzero $\varepsilon_\mathrm{ff}$, there exists a sufficiently large noise temperature $T_\mathrm{ff}$ that saturates the Holevo quantity.

\hyperref[Fig6:secureFidelity]{Figure\,6}(b) shows the corresponding teleportation fidelities for the same operating parameters as in \hyperref[Fig6:secureFidelity]{Figure\,6}(a). We find that for sufficiently large $G \gg 1$, the secure fidelity $F_\mathrm{s}$ is independent of $\varepsilon_\mathrm{ff}$. This is because Eve can always saturate the Holevo quantity, which is only limited by the pre-shared entanglement squeezing $S$. \hyperref[Fig6:secureFidelity]{Figure\,6}(c) shows $F_\mathrm{s}$ as a function of the measurement gain $G$ and resource squeezing $S$. We see that $F_\mathrm{s}$ moderately decreases for greater $G$, as the Bell-type measurement becomes more projective, and drastically decreases for greater $S$, as the private quantum correlations become stronger. In the projective measurement limit $G \to \infty$, a minimum TMS squeezing level of $S = \SI{2.39}{\decibel}$ is required to achieve security. At low gain values, secure communication is possible for lower squeezing levels because the protocol is no longer genuine quantum teleportation, but rather resembles direct quantum state transfer through the feedforward channel. Here, security is instead derived from a vanishingly small noise temperature $T_\mathrm{ff}$ of the environmental bath, and Eve's correspondingly small TMS resource state in the entangling cloner attack. Therefore, the relevant parameter regime for genuine quantum teleportation is only at squeezing levels $S \ge \SI{2.39}{\decibel}$ and relatively large gains $G \ge \SI{20}{\decibel}$.

In the limiting case of $G \to \infty$ and $G \gg W_\mathrm{ff}$, where the quantum teleportation protocol completely corrects for the feedforward channel imperfections, we can obtain an analytical expression for the mutual information
\begin{equation}\label{Eq:LimitingBob}
    I(\mathrm{A}:\mathrm{B}) = \ln\left( 1 + \frac{4 \sigma^2}{1 + 2e^{-2r}} \right).
\end{equation}
Furthermore, if we take $T_\mathrm{ff} \gg 1$ and $G \gg W_\mathrm{ff}/\varepsilon_\mathrm{ff}$, where Eve has complete information about the feedforward signal, we can obtain the Holevo quantity
\begin{equation}\label{Eq:LimitingEve}
    \chi_\mathrm{E} = \ln\left( 1 + \frac{4 \sigma^2}{1 + \cosh(2r)} \right).
\end{equation}
We then compare equations\,(\ref{Eq:LimitingBob}) and (\ref{Eq:LimitingEve}) to find that they are equal when $r = \ln(3)/4 \approx \SI{2.39}{\decibel}$. This means that for an arbitrary public channel (feedforward) eavesdropper, quantum teleportation can only be secure if the shared TMS resource has a squeezing level of $S \ge \SI{2.39}{\decibel}$, which agrees with the results of our numerical simulations in \hyperref[Fig6:secureFidelity]{Figure\,6}(c). Similarly, the secure fidelity in the limit $G \to \infty$ is given by
\begin{equation}\label{Eq:LimitingF}
    F_\mathrm{s} = \frac{2}{2+\cosh(2r)},
\end{equation}
which also coincides with the numerical results. From equation\,(\ref{Eq:LimitingF}), we have $F_\mathrm{s} = 2/3$ at $S = \SI{0}{\decibel}$, reproducing the asymptotic no-cloning threshold. However, in experiment, the analog quantum teleportation protocol requires a TMS resource with squeezing below half the vacuum level, $S \ge 10\log_{10}(2) \approx \SI{3}{\decibel}$, in order to produce a teleportation fidelity of $F = 2/3$. Thus, the numerical simulation only matches the theory results for $S \ge \SI{2.39}{\decibel}$.

Finally, we consider security against the public channel eavesdropper for truncated codebooks. For genuine quantum teleportation with $G \gg 1$, the feedforward signal can be treated classically. Obtaining secure communication requires that $I(\mathrm{A}:\mathrm{B}) > \chi_\mathrm{E}$. This condition only depends on the signal-to-noise ratios of transmitted code states and is independent of the codebook shape. The codebook distribution merely affects the nominal information content due to the varying amount of available code states. Thus, we find that the $F_\mathrm{s}$ threshold derived from the Gaussian codebook is equivalent to the secure fidelity for the truncated Gaussian and uniform codebooks. More details about the calculation of security thresholds for those codebooks are provided in \hyperref[Appendix:MutualInformationHolevo]{Appendix\,B}.

\section{Conclusion}
\label{Section:Conclusion}

In this article, we have examined an analog CV quantum teleportation protocol, interpreting the public feedforward channel as an information carrier and the quantum entanglement resource as a cipher. We have also discussed the analog quantum teleportation as an error-correcting protocol for feedforward imperfections in the limit of a large measurement gain $G \to \infty$ and vanishing coupling $\eta \to 0$, as well as under a perfect displacement-matching condition $G\eta (1-\varepsilon_\mathrm{ff}) = 4$. We have shown that quantum teleportation is resilient against arbitrary losses and noise in the feedforward channel. However, losses in the entanglement distribution channel degrade teleportation fidelity and lead to constraints on the optimal resource squeezing level $S$. We have considered state-of-the-art experimental parameters from microwave teleportation experiments and demonstrated that it is already possible to reach a teleported state fidelity beyond the classical threshold $F_\mathrm{cl} = 1/2$ at liquid helium temperatures even with significant feedforward losses $\varepsilon_{\mathrm{ff}} \simeq \SI{6}{\decibel}$. Furthermore, we have considered practical finite-energy codebooks corresponding to truncated uniform and truncated Gaussian distributions, which are experimentally more relevant than the ideal Gaussian distributions with infinite tails. For these truncated codebooks, we have found no-cloning fidelities that are higher than the asymptotic no-cloning fidelity $F_\mathrm{nc} = 2/3$, which traditionally is considered an important benchmark in quantum communication with CV states. We have also considered a more restricted, but practically relevant, scenario of a public channel eavesdropper. For this scenario, we have estimated the secure fidelity $F_\mathrm{s}$, where only the feedforward communication channel in quantum teleportation is subject to eavesdropping. This fidelity varies over a wide range as a function of the resource squeezing $S$, and provides a valuable benchmark for evaluating public channel security in quantum teleportation protocols.

Overall, our results demonstrate the growing relevance of microwave quantum communication in practical scenarios, where channel losses and noise are far from being negligible. The quantum teleportation protocol plays a particular role in such scenarios, since it is able to correct for some of these imperfections and provides a route towards the secure and high-fidelity distribution of fragile quantum states between microwave quantum nodes. Our results are highly relevant to the emerging microwave quantum networks, which are considered among the most promising candidates for interconnecting remote superconducting quantum information processing nodes and, thus, providing a backbone for future scalable superconducting quantum architectures. Finally, our results also indicate the feasibility of microwave quantum communication over short distances at room temperature conditions~\cite{Fesquet2023,Fesquet2024}, which may find useful applications in high-frequency technologies and beyond.


\appendix
\newcounter{appendix}
\makeatletter
\@addtoreset{figure}{appendix}
\makeatother
\renewcommand\thefigure{\thesection-\arabic{figure}}

\section{Theory model of analog CV quantum teleportation}
\label{Appendix:CoherentStateTeleportation}
\setcounter{figure}{0}
\stepcounter{appendix}

\begin{figure*}[t]
	\centering
	\includegraphics[width=\linewidth]{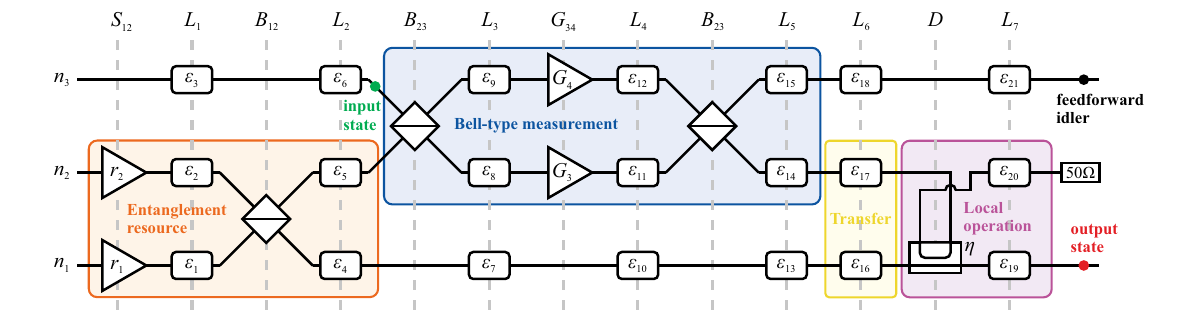}
	\caption{Full theory model for analog continuous-variable quantum teleportation with losses and added noise.}
	\label{AppendixFig1:theoryModel}
\end{figure*}

The analog CV quantum teleportation of coherent states involves only Gaussian states and Gaussian operations. Therefore, we can use the covariance matrix formalism to fully describe this protocol~\cite{Weedbrook2012}. In the following, we use the vacuum variance definition of $1/4$ per field quadrature. For the matrix representations, we use the basis $\{\ket{p_1}, \ket{q_1}, \ket{p_2}, \ket{q_2}, \ket{p_3}, \ket{q_3}\}$ where $\ket{p_i}$, $\ket{q_i}$ are the $p$- and $q$-quadrature components of mode $i$. We use $\mathbb{1}_2$ for the $2 \times 2$ identity matrix, $\mathbb{0}_2$ for the $2 \times 2$ zero matrix, and $\sigma_z$ for the Pauli-$z$ matrix. \hyperref[AppendixFig1:theoryModel]{Figure\,A-1} illustrates our full theory model of analog quantum teleportation, which includes all operators and path losses. We note that $\varepsilon_\mathrm{ent}$ and $\varepsilon_\mathrm{ff}$ in the article correspond to $\varepsilon_{16}$ and $\varepsilon_{17}$, respectively.

The initial displacement vector and covariance matrix for Alice and Bob are given by
\begin{align}
	d_\mathrm{initial} &= (0, 0, 0, 0, \Re(\alpha), \Im(\alpha))^\intercal, \\
	V_\mathrm{initial} &= \frac{1}{4}
	\begin{pmatrix}
		(1+2 n_1) \mathbb{1}_2 & \mathbb{0}_2 & \mathbb{0}_2 \\
		\mathbb{0}_2 & (1+2 n_2) \mathbb{1}_2 & \mathbb{0}_2 \\
		\mathbb{0}_2 & \mathbb{0}_2 & (1+2 n_3) \mathbb{1}_2
	\end{pmatrix},
\end{align}
where $\alpha$ is the complex displacement amplitude and $n_1$, $n_2$, $n_3$ are the input noise photon numbers at each mode, respectively. At the start of the teleportation protocol, the first mode contains Bob's part of the TMS state, the second mode contains Alice's part of the TMS state, and the third mode contains the input coherent state. To describe the input state ensemble, when using a zero-mean Gaussian codebook with variance $\sigma^2$, we take $\alpha = 0$ and replace the third mode of $V_\mathrm{initial}$ with  $(4\sigma^2 + 1+2 n_3) \mathbb{1}_2$.

The full teleportation protocol is encoded by~\cite{Fedorov2021,Yam2025}
\begin{equation}\label{Eq:Teleportation}
	\hat{T} = \hat{L}_7 \hat{D} \hat{L}_6 \hat{L}_5 \hat{B}_{23} \hat{L}_4 \hat{G}_{34} \hat{L}_3 \hat{B}_{23} \hat{L}_2 \hat{B}_{12} \hat{L}_1 \hat{S}_{12}.
\end{equation}
We write the squeezing operator as
\begin{equation}
	\hat{S}_{12} = \hat{R}_{12} \hat{J}_{12} \hat{R}_{12}^{\dagger},
\end{equation}
with
\begin{align}
	\hat{J}_{12} &=
	\begin{pmatrix}
		e^{-r_1} & 0 & 0 & 0 & 0 & 0 \\
		0 & e^{r_1} & 0 & 0 & 0 & 0 \\
		0 & 0 & e^{-r_2} & 0 & 0 & 0 \\
		0 & 0 & 0 & e^{r_2} & 0 & 0 \\
		0 & 0 & 0 & 0 & 1 & 0 \\
		0 & 0 & 0 & 0 & 0 & 1
	\end{pmatrix}, \\
	\hat{R}_{12} &=
	\begin{pmatrix}
		\cos{\gamma_1} & -\sin{\gamma_1} & 0 & 0 & 0 & 0 \\
		\sin{\gamma_1} & \cos{\gamma_1} & 0 & 0 & 0 & 0 \\
		0 & 0 & \cos{\gamma_2} & -\sin{\gamma_2} & 0 & 0 \\
		0 & 0 & \sin{\gamma_2} & \cos{\gamma_2} & 0 & 0 \\
		0 & 0 & 0 & 0 & 1 & 0 \\
		0 & 0 & 0 & 0 & 0 & 1 \\
	\end{pmatrix}.
\end{align}
We set the squeezing factors $r_1 = r_2 = r$ and the squeezing angles $\gamma_1 = 0$, $\gamma_2 = \pi/2$ for the TMS resource. The beam splitter operators are expressed as
\begin{align}
	\hat{B}_{12} &= \frac{1}{\sqrt{2}}
	\begin{pmatrix}
		\mathbb{1}_2 & \mathbb{1}_2 & \mathbb{0}_2 \\
		-\mathbb{1}_2 & \mathbb{1}_2 & \mathbb{0}_2 \\
		\mathbb{0}_2 & \mathbb{0}_2 & \sqrt{2} \mathbb{1}_2
	\end{pmatrix}, \\
	\hat{B}_{23} &= \frac{1}{\sqrt{2}}
	\begin{pmatrix}
		\sqrt{2} \mathbb{1}_2 & \mathbb{0}_2 & \mathbb{0}_2 \\
		\mathbb{0}_2 & \mathbb{1}_2 & \mathbb{1}_2 \\
		\mathbb{0}_2 & -\mathbb{1}_2 & \mathbb{1}_2
	\end{pmatrix}.
\end{align}
The quadrature measurement (strong phase-sensitive amplification) operator can be written as
\begin{equation}
	\hat{G}_{34} = \hat{R}_{34} \hat{J}_{34} \hat{R}_{34}^{\dagger},
\end{equation}
with
\begin{align}
	\hat{J}_{34} &=
	\begin{pmatrix}
		1 & 0 & 0 & 0 & 0 & 0 \\
		0 & 1 & 0 & 0 & 0 & 0 \\
		0 & 0 & 1/\sqrt{G_3} & 0 & 0 & 0 \\
		0 & 0 & 0 & \sqrt{G_3} & 0 & 0 \\
		0 & 0 & 0 & 0 & 1/\sqrt{G_4} & 0 \\
		0 & 0 & 0 & 0 & 0 & \sqrt{G_4}
	\end{pmatrix}, \\
	\hat{R}_{34} &=
	\begin{pmatrix}
		1 & 0 & 0 & 0 & 0 & 0 \\
		0 & 1 & 0 & 0 & 0 & 0 \\
		0 & 0 & \cos{\gamma_3} & -\sin{\gamma_3} & 0 & 0 \\
		0 & 0 & \sin{\gamma_3} & \cos{\gamma_3} & 0 & 0 \\
		0 & 0 & 0 & 0 & \cos{\gamma_4} & -\sin{\gamma_4} \\
		0 & 0 & 0 & 0 & \sin{\gamma_4} & \cos{\gamma_4} \\
	\end{pmatrix}.
\end{align}
We set the gains $G_3 = G_4 = G$ and the measurement angles $\gamma_3 = \gamma_1$, $\gamma_4 = \gamma_2$ for the Josephson interferometer. In order to genuinely implement analog quantum teleportation, the measurement gain must be large, $G \gg 1$. We write the directional coupler operator
\begin{equation}
	\hat{D} =
	\begin{pmatrix}
		\sqrt{1-\eta} \mathbb{1}_2 & \sqrt{\eta} \mathbb{1}_2 & \mathbb{0}_2 \\
		-\sqrt{\eta} \mathbb{1}_2 & \sqrt{1-\eta} \mathbb{1}_2 & \mathbb{0}_2 \\
		\mathbb{0}_2 & \mathbb{0}_2 & \mathbb{1}_2 \\
	\end{pmatrix},
\end{equation}
where $\eta$ is the coupling factor of the directional coupler. We express the loss operators for each segment of the experimental setup as
\begin{equation}
	\hat{L}_{i+1} =
	\begin{pmatrix}
		\sqrt{1-\varepsilon_{3i+1}} \mathbb{1}_2 & \mathbb{0}_2 & \mathbb{0}_2 \\
		\mathbb{0}_2 & \sqrt{1-\varepsilon_{3i+2}} \mathbb{1}_2 & \mathbb{0}_2 \\
		\mathbb{0}_2 & \mathbb{0}_2 & \sqrt{1-\varepsilon_{3i+3}} \mathbb{1}_2 \\
	\end{pmatrix}.
\end{equation}
During the teleportation procedure, noise from the local thermal baths is coupled into the propagating quantum states via the local losses $\varepsilon_j$, which can be modeled using the beam splitter operation with transmissivity $\varepsilon_j$. The final displacement vector and covariance matrix are given by
\begin{align}
	d_\mathrm{final} &= \hat{T} d_\mathrm{initial}, \\
	V_\mathrm{final} &= \hat{T} V_\mathrm{initial} \hat{T}^{\dagger} + \mathbb{N}, \label{Eq:TeleportationVariance}
\end{align}
where the matrix $\mathbb{N}$ represents the coupled noise due to losses in the protocol. We can simply extract the first modes of $d_\mathrm{final}$ and $V_\mathrm{final}$ in order to obtain Bob's output state at the end of the teleportation procedure. To perform a realistic simulation of this teleportation protocol, we use experimental parameters from Ref.\,\cite{Yam2025}.

In order to model the eavesdropper (Eve) attack on the public communication channel, we consider the teleportation protocol at the transfer step (denoted by the yellow box in \hyperref[AppendixFig1:theoryModel]{Figure\,A-1})
\begin{equation}
	\hat{T}_\mathrm{trans} = \hat{L}_5 \hat{B}_{23} \hat{L}_4 \hat{G}_{34} \hat{L}_3 \hat{B}_{23} \hat{L}_2 \hat{B}_{12} \hat{L}_1 \hat{S}_{12}.
\end{equation}
The displacement vector and covariance matrix at the transfer step are given by
\begin{align}
	d_\mathrm{trans} &= \hat{T}_\mathrm{trans} d_\mathrm{initial}, \\
	V_\mathrm{trans} &= \hat{T}_\mathrm{trans} V_\mathrm{initial} \hat{T}_\mathrm{trans}^{\dagger} + \mathbb{N}_\mathrm{trans},
\end{align}
where the matrix $\mathbb{N}_\mathrm{trans}$ represents the coupled noise up to the transfer step. We can obtain the feedforward signal state, described using $d_\mathrm{ff}$ and $V_\mathrm{ff}$, by extracting the second modes of $d_\mathrm{trans}$ and $V_\mathrm{trans}$, respectively. The displacement vector and covariance matrix for Eve's attack are then given by
\begin{align}
	d_\mathrm{attack} &= (d_\mathrm{ff}, 0, 0, 0, 0)^\intercal, \\
	V_\mathrm{attack} &= \frac{1}{4}
	\begin{pmatrix}
		V_\mathrm{ff} & \mathbb{0}_2 & \mathbb{0}_2 \\
        \mathbb{0}_2 & W_\mathrm{ff} \mathbb{1}_2 & \sqrt{W_\mathrm{ff}^2 - 1} \sigma_z \\
		\mathbb{0}_2 & \sqrt{W_\mathrm{ff}^2 - 1} \sigma_z & W_\mathrm{ff} \mathbb{1}_2
	\end{pmatrix},
\end{align}
where $W_\mathrm{ff}$ characterizes the power of Eve's TMS resource in the context of the entangling cloner attack~\cite{Braunstein2001a}. The first mode contains the feedforward signal state ensemble, and the second and third modes contain Eve's entanglement resource. To siphon the feedforward signal, Eve couples to the feedforward channel via the operation
\begin{equation}
	\hat{E} =
	\begin{pmatrix}
		\sqrt{1-\varepsilon_\mathrm{ff}} \mathbb{1}_2 & \sqrt{\varepsilon_\mathrm{ff}} \mathbb{1}_2 & \mathbb{0}_2 \\
		-\sqrt{\varepsilon_\mathrm{ff}} \mathbb{1}_2 & \sqrt{1-\varepsilon_\mathrm{ff}} \mathbb{1}_2 & \mathbb{0}_2 \\
		\mathbb{0}_2 & \mathbb{0}_2 & \mathbb{1}_2
	\end{pmatrix},
\end{equation}
where $\varepsilon_\mathrm{ff}$ is the feedforward losses. The resulting displacement vector and covariance matrix are given by
\begin{align}
	d_\mathrm{attack}' &= \hat{E} d_\mathrm{attack}, \\
	V_\mathrm{attack}' &= \hat{E} V_\mathrm{attack} \hat{E}^{\dagger},
\end{align}
from which we can extract the second and third modes to obtain Eve's output state after the attack.

\section{Calculation of mutual information and Holevo quantity}
\label{Appendix:MutualInformationHolevo}
\setcounter{figure}{0}
\stepcounter{appendix}

Information shared between Alice and Bob is quantified by the mutual information
\begin{equation}
	I(\mathrm{A}:\mathrm{B}) = H(\hat{\rho}_\mathrm{A}) + H(\hat{\rho}_\mathrm{B}) - H(\hat{\rho}_\mathrm{AB}),
\end{equation}
where $H(\cdot)$ is the von Neumann entropy and $\hat{\rho}_\mathrm{A}$, $\hat{\rho}_\mathrm{B}$, and $\hat{\rho}_\mathrm{AB}$ are the density matrices of the Alice subsystem, Bob subsystem, and joint system, respectively. For the analog CV quantum teleportation protocol of coherent states, we can calculate the mutual information using~\cite{Braunstein2005}
\begin{equation}\label{AppendixEq:MI}
	I(\mathrm{A}:\mathrm{B}) = \iint d\alpha d\beta~ P_\mathrm{cond}(\beta|\alpha) P_\mathrm{in}(\alpha) \ln\left( \frac{ P_\mathrm{cond}(\beta|\alpha) }{ P_\mathrm{out}(\beta) } \right),
\end{equation}
where $\alpha$ and $\beta$ are the complex displacement amplitudes of the input and teleported output states, respectively. The integration is done over the two-dimensional complex planes of $\alpha$ and $\beta$. The input state ensemble distributions $P_\mathrm{in}(\alpha)$ considered in this paper are given by equations\,(\ref{Eq:gaussian}), (\ref{Eq:truncatedUniform}), and (\ref{Eq:truncatedGaussian}). The conditional distribution is given by
\begin{equation}
	P_\mathrm{cond}(\beta|\alpha) = \frac{1}{2\pi v_\mathrm{out}} e^{-\frac{|\beta - \sqrt{k}\alpha|^2}{2 v_\mathrm{out}}},
\end{equation}
where $k = G\eta(1-\varepsilon_\mathrm{ff})/4$ is the overall amplification and $v_\mathrm{out}$ is the teleported state variance (defined in equation\,(\ref{Eq:outputCovariance})). The teleported output state ensemble distribution is given by
\begin{equation}
	P_\mathrm{out}(\beta) = \int d\alpha~ P_\mathrm{cond}(\beta|\alpha) P_\mathrm{in}(\alpha).
\end{equation}
For the Gaussian codebook with variance $\sigma^2$, we can analytically calculate the teleported output state ensemble distribution
\begin{equation}
	P_\mathrm{out}(\beta) = \frac{1}{2\pi (v_\mathrm{out} + k\sigma^2)} e^{-\frac{|\beta|^2}{2 (v_\mathrm{out} + k\sigma^2)}},
\end{equation}
and, hence, obtain its corresponding mutual information
\begin{equation}\label{AppendixEq:BobMI}
	I(\mathrm{A}:\mathrm{B}) = \ln\left( 1 + \frac{ k\sigma^2 }{ v_\mathrm{out} } \right).
\end{equation}
For the truncated Gaussian codebook with variance $\sigma^2$ and cutoff photon number $N$, the teleported output state ensemble distribution is given by
\begin{equation}
	\begin{split}
		& P_\mathrm{out}(\beta) \\
        &= \frac{1}{2\pi (v_\mathrm{out} + k\sigma^2) (1 - e^{-N/2\sigma^2})} \exp\left[ \frac{-|\beta|^2}{2 (v_\mathrm{out} + k\sigma^2)} \right] \\
		&\quad\times \biggl\{ 1 - \exp\left[ -\zeta N - \xi|\beta|^2 \right] \Phi_3 \left( 1,1; \xi|\beta|^2, \xi\zeta N |\beta|^2 \right) \biggr\},
	\end{split}
\end{equation}
\begin{equation}
	\zeta = \frac{1}{2\sigma^2} + \frac{k}{2v_\mathrm{out}},
    \quad\quad
	\xi = \frac{1}{\zeta} \frac{k}{4v_\mathrm{out}^2},
\end{equation}
\begin{equation}
	\Phi_3(a,b;x,y) = \sum_{m=0}^{\infty} \sum_{n=0}^{\infty} \frac{(a)_{m}}{m!n! (b)_{m+n}} x^m y^n,
\end{equation}
where $\Phi_3(a,b;x,y)$ is the third Humbert hypergeometric function.
The distribution for the truncated uniform codebook can be obtained by taking $\sigma \to \infty$ and $\lim_{\sigma \to \infty} 2\sigma^2 (1-e^{-N/2\sigma^2}) = N$. The corresponding mutual information values are computed numerically using equation\,(\ref{AppendixEq:MI}).

\begin{figure}[t]
	\centering
	\includegraphics[width=\linewidth]{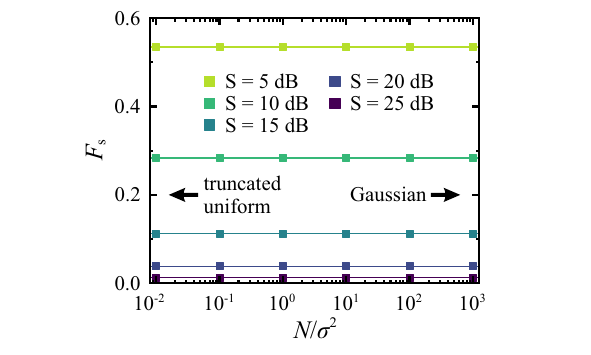}
	\caption{Secure fidelity $F_\mathrm{s}$ for the truncated Gaussian codebook as a function of the truncation-to-variance ratio $N/\sigma^2$. Square markers represent numerically calculated $F_\mathrm{s}$ for the truncated Gaussian codebook. Solid lines represent $F_\mathrm{s}$ for the Gaussian codebook derived from equation\,(\ref{Eq:LimitingF}). The values approach the truncated uniform and Gaussian codebook results when $N/\sigma^2 \to 0$ and $N/\sigma^2 \to \infty$, respectively.}
	\label{AppendixFig4:truncatedSecurity}
\end{figure}

An upper bound for Eve's accessible information is given by the Holevo quantity~\cite{Holevo1973}
\begin{equation}
	\chi_\mathrm{E} = H(\hat{\rho}_\mathrm{E}) - \int d\alpha~ P_\mathrm{in}(\alpha) H(\hat{\rho}_\alpha),
\end{equation}
where $\hat{\rho}_\mathrm{E} = \int d\alpha~ P_\mathrm{in}(\alpha) \hat{\rho}_\alpha$ is the state ensemble at Eve and $\hat{\rho}_\alpha$ are the individual states at Eve corresponding to \emph{a priori} probability distribution $P_\mathrm{in}(\alpha)$. The state ensemble at Eve is formed by superimposing the input state ensemble with one part of the TMS resource state via a balanced beam splitter, so its distribution can be described by the convolution
\begin{equation}
	P_\mathrm{E}(\beta) = \int d\alpha~ P_\mathrm{th}(\beta-\alpha) P_\mathrm{in}(\alpha),
\end{equation}
where we have the thermal state distribution
\begin{equation}
	P_\mathrm{th}(\alpha) = \frac{1}{2\pi v_\mathrm{eve}} e^{-|\alpha|^2/2 v_\mathrm{eve}},
\end{equation}
and $v_\mathrm{eve} = [1+\cosh(2r)]/4$. Here, we neglect the feedforward amplification $G/4$ because a rescaling does not affect information content. Since the individual thermal coherent states $\hat{\rho}_\alpha$ have identical entropy as the zero-mean thermal state, we can calculate the Holevo quantity using
\begin{equation}
\begin{split}
	\chi_\mathrm{E} &= H(\hat{\rho}_\mathrm{E}) - H(\hat{\rho}_\alpha) \\
    &= - \int d\beta P_\mathrm{E}(\beta) \ln P_\mathrm{E}(\beta) + \int d\alpha P_\mathrm{th}(\alpha) \ln P_\mathrm{th}(\alpha).
\end{split}
\end{equation}
For the Gaussian codebook with variance $\sigma^2$, we can obtain an analytical expression for the Holevo quantity
\begin{equation}\label{AppendixEq:Holevo}
    \chi_\mathrm{E} = \ln\left( 1 + \frac{ 4\sigma^2 }{ 1+\cosh(2r) } \right).
\end{equation}
For the truncated uniform and truncated Gaussian codebooks, this computation is done numerically.

We numerically calculate $I(\mathrm{A}:\mathrm{B})$ and $\chi_\mathrm{E}$ for various truncated codebook parameters to determine the secure fidelity $F_\mathrm{s}$. \hyperref[AppendixFig4:truncatedSecurity]{Figure\,B-1} shows $F_\mathrm{s}$ as a function of the truncation-to-variance ratio $N/\sigma^2$ for a truncated Gaussian codebook with \emph{a priori} probability distribution $P(\alpha,\sigma,N)$ (defined in equation\,(\ref{Eq:truncatedGaussian})). We observe that the truncated Gaussian codebook $F_\mathrm{s}$ values match the Gaussian codebook result for all $N/\sigma^2$. This is because secure communication only depends on the relative information between $I(\mathrm{A}:\mathrm{B})$ and $\chi_\mathrm{E}$, which only depends on the signal-to-noise ratios of transmitted code states and is independent of the codebook shape. Thus, the Gaussian codebook result is representative for any codebook.


\acknowledgments

We acknowledge helpful discussions with Roberto Di Candia related to the finite-energy codebooks and with Florian Fesquet related to the security analysis. We acknowledge support by the German Research Foundation via Germany's Excellence Strategy (EXC-2111-390814868), the EU Quantum Flagship project QMiCS (Grant No.~820505), and the German Federal Ministry of Education and Research via the project QUARATE (Grant No. 13N15380). This research is part of the Munich Quantum Valley, which is supported by the Bavarian state government with funds from the Hightech Agenda Bayern Plus.


\bibliography{Bibliography}

\end{document}